\documentclass[fleqn,twoside]{article}
\usepackage{espcrc2}

\usepackage{graphicx}
\usepackage[figuresright]{rotating}

\newcommand{\AmS}{{\protect\the\textfont2
  A\kern-.1667em\lower.5ex\hbox{M}\kern-.125emS}}
\def\amuhad{$a_\mu^{\rm hvp}$}
\newcommand{\bda}{\begin{\displaymath}\begin{array}{rl}}
\newcommand{\eda}{\end{array}\end{displaymath}}
\newcommand{\be}{\begin{equation}}
\newcommand{\ee}{\end{equation}}
\newcommand{\bdm}{\begin{displaymath}}
\newcommand{\edm}{\end{displaymath}}
\newcommand{\bea}{\begin{eqnarray}}
\newcommand{\eea}{\end{eqnarray}}

\newcommand{\fs}{\,.}
\newcommand{\co}{\,,}

\title{Hadronic contributions to $a_\mu$ below one GeV}

\author{Gilberto Colangelo
\address{Institute f\"ur Theoretische Physik
        der Universit\"at Bern\\
        Sidlerstr. 5 3012 Bern Switzerland}
\thanks{This work is 
partly supported by the Swiss National Science
Foundation and by RTN, BBW-Contract No. 01.0357 and EC-Contract
HPRN--CT2002--00311 (EURIDICE)}%
}
       
\begin{document}

\begin{abstract}
I present a method for evaluating the hadronic vacuum polarization
contribution below 1 GeV to $a_\mu$ which relies on analyticity, unitarity
and chiral symmetry, as well as on data. The main advantage is that in the
region just above threshold, where data are either scarce or have large
errors, these theoretical constraints are particularly strong, and
therefore allow us to reduce the uncertainties with respect to a purely
data--based evaluation.  Some preliminary numerical results are presented
as illustration of the method. \vspace{1pc}
\end{abstract}

\maketitle

\section{Introduction}
The anomalous magnetic moment of the muon $a_\mu$ is now known
experimentally to an extremely high precision \cite{Brown:2001mg}, and can
therefore be used as a thorough test of the standard model. The most
uncertain and debated part of the standard model calculation of $a_\mu$ is
the hadronic vacuum polarization contribution \amuhad\ . In order to make
this test of the standard model a significant one, we need to make
substantial progress in the evaluation of the hadronic contributions, and
in particular of the leading one, \amuhad\ -- which was precisely the
scope of this Workshop. This contribution can be calculated
in terms of another experimentally measured quantity, the cross section
$e^+ e^- \to$ hadrons, and any improvement in the measurement of this cross
section is immediately reflected in \amuhad\ . Indeed many discussions at
the Workshop concerned different possible ways to measure the $e^+e^-$
hadronic cross section, and have given us an overview of an impressive
amount of experimental and theoretical work devoted to this problem. Given
a set of data points at different center of mass energies for $\sigma(e^+
e^- \to \mbox{hadrons})$, the evaluation of \amuhad\ amounts to the
calculation of an integral of a function once one knows it at a discrete
set of points.  This problem is standard, and can be solved in different
ways -- the simplest of them being the use of the trapezoidal
rule, as done, e.g. in Refs.~\cite{EJ,DEHZ-1}.

The use of the trapezoidal rule, or variants thereof, reduces the role of
theory to an absolute minimum -- the evaluation of \amuhad\ is done almost
only with data. The role of theory cannot be reduced to zero because it is
needed in the two extreme regions $s \sim 4 M_\pi^2$ and $s \to \infty$. In
the latter region one can make use of perturbative QCD, and in the former
one of chiral perturbation theory (CHPT). Actually, CHPT cannot give a
sharp numerical prediction for the pion vector form factor $F_V(s)$ (the
dominating contribution below 1 GeV) close to threshold: it only predicts
that the behaviour is very smooth, well approximated by a polynomial, whose
coefficients are related to a few of the low energy constants of the chiral
Lagrangian. The use of CHPT in this region amounts to an extrapolation of
the data at somewhat higher energies down to threshold with a polynomial of
low degree \cite{EJ,DEHZ-1}.

My aim here is to show that not only close to threshold, but even up to 1
GeV the use of some theory is actually quite useful, especially because it
allows one to reduce the uncertainty in \amuhad\ -- and this region
contributes the largest fraction of the total error. Theory in this case
means some very general properties which we know the vector form factor
$F_V(s)$ must satisfy: analyticity and unitarity.  Combining these
properties with chiral symmetry, which is relevant in the very low energy
region, one can construct a representation which is very constraining for
the form factor: the remaining little freedom can be fixed with the help of
data as I illustrate in what follows.

\section{Definition of \amuhad\ }
\begin{figure}[thb]
\begin{center}
\leavevmode
\includegraphics[width=6cm]{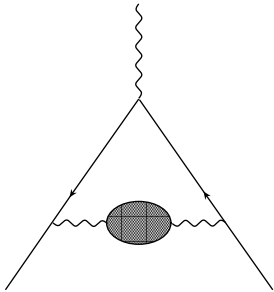}
\end{center}
\caption{\label{fig:hvp} Hadronic vacuum polarization contribution to
  $(g-2)_\mu$.} 
\end{figure}

The leading hadronic contribution to $(g-2)_\mu$ is due to the hadronic
vacuum polarization correction shown in Fig.~\ref{fig:hvp}. 

This contribution to $(g-2)_\mu$ is of order $\alpha^2$ and
can be expressed in terms of the cross section $\sigma(e^+ e^- \to
\mbox{hadrons})$ evaluated to leading order in $\alpha$ \cite{GR}:
\be
\label{eq:amuhvp}
a_\mu^{\mbox{\tiny hvp}}=
\left({\alpha m_\mu \over 3 \pi} \right)^2 \int_{4 M_\pi^2}^\infty ds {
  \hat K(s) R^{(0)}(s) \over s^2} \co
\ee
where $\hat K(s)$ is a known kernel, see e.g. \cite{Jegerlehner}, and 
\be
R^{(0)}(s) \equiv {\sigma^{(0)}(e^+ e^- \to \mbox{hadrons}) \over 4\pi
  \alpha^2/3s} \co
\ee
where the superscript indicates that the cross section has to be
taken to leading order in $\alpha$\footnote{We do not discuss here the
  problem of extracting the leading order term from data.}. At low energy
$\sqrt{s} \leq 1$ GeV, the contribution of the two--pion state dominates
the cross section 
\be
R^{(0)} (s) \simeq R^{\pi\pi}(s)={1 \over 4} \left(1-4M_\pi^2/s \right)^{3/2}
\left|F_V(s) \right|^2 \fs
\ee
The latter, as shown, is given by the vector form factor of the pion (again
to leading order in $\alpha$), which is the quantity we will now discuss.

\section{The pion vector form factor}
The pion vector form factor $F_V(s)$ is an analytic function of $s$ in the
whole complex plane, with the exception of a cut on the real axis for $s
\geq 4 M_\pi^2$: approaching the real axis from above the form factor stays
complex and can be described in terms of two real functions, its modulus
and phase: 
\be
F_V(s)=| F_V(s)| e^{i\delta(s)} \fs
\ee
Omn\`es \cite{Omnes} has shown that analyticity relates the modulus and the
phase, such that the whole function can be given in closed form in terms of
its phase:
\be
F_V(s)= P(s) \exp \left[
  \frac{s}{\pi}\int_{4M_\pi^2}^\infty dx\,\frac{\delta(x)} 
{x\,(x-s)}  \right] \co
\label{eq:omnes}
\ee
where $P(s)$ is a polynomial which determines the behaviour of the function
at infinity, or number and position of its zeros. In order to respect
charge conservation $F_V(0)=1$, we set $P(0)=1$. The representation
(\ref{eq:omnes}) makes it apparent that if one knows the phase on the cut
and the zeros of the form factor one can calculate the form factor
everywhere in the complex plane.

In the elastic region Watson's theorem relates the phase of the vector form
factor to the phase of the $\pi \pi$ scattering amplitude with the same
quantum numbers, $I=\ell=1$:
\be
\delta(s)=\delta_1^1(s) \quad \mbox{for} \quad s \leq s_{in}=16 M_\pi^2 \fs
\ee
While the inelastic threshold $\sqrt{s_{in}}=.56$ GeV appears to be rather
low, inelastic contributions in this channel are known to become relevant
only above the $K \bar K$ threshold. To an excellent approximation the
phase of the vector form factor coincides with the $\pi \pi$ phase shift
$\delta_1^1$ up to 1 GeV. The latter has been recently studied in
the framework of Roy equations, first with no extra input \cite{ACGL}, and
also in combination with chiral symmetry \cite{CGL}. The upshot of these
analyses is that $\delta_1^1$ is constrained to a remarkable 
degree of accuracy up to about 0.8 GeV. In our approach we make explicit
use of these general properties of the vector form factor, and of our
knowledge of the $\pi \pi$ phase shifts. Our strategy can be summarized in
the following points:
\begin{enumerate}
\item we construct a convenient representation which automatically respects
  the properties of analyticity, unitarity and chiral symmetry;
\item we fix the free parameters which appear in this representation by
  fitting data;
\item we evaluate the integral (\ref{eq:amuhvp}) up to 1 GeV using our
  analytic representation of the form factor.
\end{enumerate}
A similar strategy was also adopted in Ref.~\cite{TY}, but with a limited
use of chiral symmetry.

\section{A convenient representation of the vector form factor}
As discussed above, the vector form factor in the low energy region is to a
large extent dominated by the $\pi \pi$ phase shift $\delta_1^1$. Inelastic
effects, which are small, but nonnegligible at the needed level of
accuracy, will also be taken into account and parametrized in terms of a
smooth function which has the correct analytic properties. In order to
evaluate \amuhad\ we need the vector form factor to leading order in
$\alpha$, i.e. with electromagnetic interactions switched off -- our form
factor $F_V$ does not include vacuum polarization corrections. To make the
connection to the $\pi \pi$ phase determined in \cite{CGL} it is also
convenient to work in the isospin limit of strong interactions\footnote{The
quark mass should be chosen such that the common pion mass is equal to the
physical charged pion mass} $m_u=m_d$. In the $\pi \pi$ scattering
amplitude these isospin-violating effects only show up at order
$(m_u-m_d)^2$ and are negligible. In the form factor, however, they are
linear in the quark mass difference, and moreover they are enhanced, in a
certain energy region, by the small mass difference between the $\rho$ and
$\omega$ mesons, which appears in the denominator. This enhanced
isospin-violating effect cannot be neglected at the level of accuracy
which we are working at. We are therefore going to represent the form factor
as a product of three functions that account for the prominent
singularities in the low energy region \cite{arkadyfest}: \be\label{three
factors} F(s)= G_1(s)\cdot G_2(s) \cdot G_\omega(s) \fs\ee

\begin{figure}[t]
\begin{center}
\includegraphics[width=7cm]{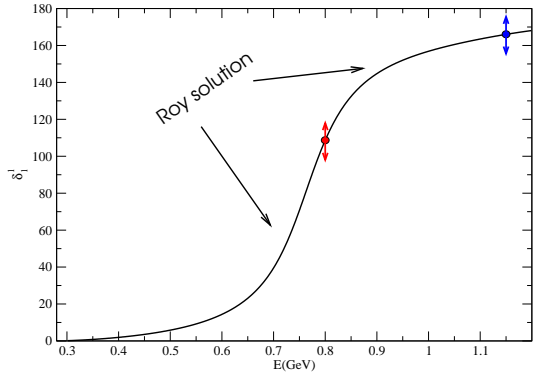}
\caption{\label{fig:d11} The $\pi \pi$ phase shift $\delta_1^1$: the value
  of the phase at the two (red and blue) dots are free and the Roy
  equations and chiral symmetry completely fix the phase everywhere else.}
\end{center}
\end{figure}

The first term is the Omn\`es factor that describes the cut due to $2\pi$
intermediate states: 
\be
G_1(s)= \exp \left\{\frac{s}{\pi}\int_{4M_\pi^2}^\infty\frac{dx\,\delta_1^1(x)}
{x\,(x-s)} \right\} \fs \ee
The phase $\delta_1^1$ which enters $G_1(s)$ is obtained with an updated
version of the analysis in \cite{CGL}: in solving Roy equations we need an
input for the imaginary part of the various partial waves at and above 0.8
GeV. In \cite{CGL} the input $I=\ell=1$ partial wave had been fixed with
the CLEO data \cite{CLEO} on the vector form factor: the most important input
parameter, the phase at $\sqrt{s_0}=0.8$ GeV had been taken equal to
\be
\delta_1^1(s_0)=(108.9 \pm 2)^\circ \fs
\label{eq:d11s0}
\ee
Above that energy we had used the parametrization of Hyams et
al. \cite{Hyams}. 
In the present work we leave $\delta_1^1(s_0)$ and $\delta_1^1(s_1)$, with
$\sqrt{s_1}=1.15$ GeV (the upper limit of validity of the Roy equations),
as free parameters and determine them by fitting data on the form factor.
Once these input parameters are given, Roy equations and chiral symmetry
fix uniquely the phase between these two points and all the way down to
threshold, as illustrated in Fig.~\ref{fig:d11}.
The phase shown in the plot corresponds to 
\be
\delta_1^1(s_0)=109.5^\circ \qquad \delta_1^1(s_1)=165.9^\circ
\ee
which are typical values we obtain in our fits, and in perfect agreement
with the input phase used in \cite{ACGL}. The error which we get on
$\delta_1^1(s_0)$ is however about a factor two smaller than that in
(\ref{eq:d11s0}). 

The function $G_\omega(s)$ contains the pole generated by $\omega$ exchange,
\bea
G_\omega(s)&=&1+\epsilon\,\frac{s}{s_\omega-s}+\ldots  \nonumber \\
s_\omega&=&(M_\omega-\mbox{$\frac{1}{2}$}\,i\,\Gamma_\omega)^2 \fs
\eea
The pole term cannot stand by itself because it fails to be real in the
spacelike region. We replace it by a dispersion integral with the proper
behaviour at threshold, but this is inessential: the representation for
$G_\omega(s)$ that we are using is 
(a) fully determined by the values of $\epsilon$, $M_\omega$ and
$\Gamma_\omega$ and 
(b) in the experimental range, $|G_\omega(s)|$ is numerically very close to
the magnitude of the pole approximation.

The function $G_2(s)$ represents the smooth background that
contains the curvature generated by the singularities not acconted for by $G_1$
and $G_\omega$. We analyze this term by means of a conformal mapping. The
$4\pi$ channel opens at $s_{in}=16\,M_\pi^2$, but phase space strongly
suppresses the strength of the corresponding branch point singularity,
which is of the form $(1-s_{in}/s)^{9/2}$. 
The transformation
\be\label{conformal map}
z=\frac{\sqrt{s_{in}-s_1}\,-\sqrt{s_{in}-s}}{\sqrt{s_{in}-s_1}\,+
\sqrt{s_{in}-s}}\ee  
maps the plane cut along $s>s_{in}$ onto the unit disk in the $z$-plane. 
It contains a free parameter $s_1$ -- the value of $s$ that gets mapped
into the origin.  We find that if $s_1$ is taken negative and sufficiently
far from the origin, the fit becomes insensitive to its specific value. We
set $s_1=-1\,\mbox{GeV}^2$.  
\begin{figure}[t]
\begin{center}
\includegraphics[width=7cm]{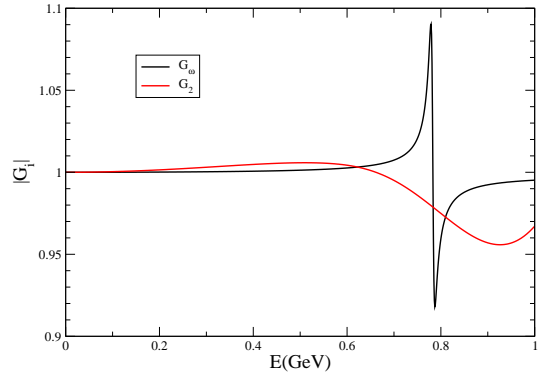}
\caption{\label{fig:G2om} Typical shapes of the factors $G_2$ and $G_\omega$
  that enter the vector form factor.}
\end{center}
\end{figure}
We approximate $G_2(s)$ by a polynomial in the variable $z$,
\be\label{G} G_2(s)=1+ \sum_{i=1}^P c_i\,(z^i-z_0^i) \co \ee
where $z_0$ is the image of the point $s=0$.
The terms involving $z_0$ ensure that the background does not modify the
charge, $G_2(0)=1$. The condition that the branch point singularity has the
form $(1-s_{in}/s)^{9/2}$ implies four constraints on the coefficients: if
we want a nontrivial contribution from $G_2(s)$ we need at least a
fifth--order polynomial. In the following we will vary $P$, the order of
the polynomial, between $0$ and $8$.
Both factors $G_2$ and $G_\omega$ are very small in the region up to 1 GeV,
as illustrated in Fig.~\ref{fig:G2om}: only $G_\omega$ generates a 10\%
effect but only in a very narrow region around $M_\omega$, otherwise both
give an effect at the few percent level. Notice that in principle $G_2$ may
have zeros somewhere in the complex $s$ plane, as allowed by the general
Omn\`es representation (\ref{eq:omnes}) -- we make no assumptions on their
position, nor on their number, if we vary $P$, and let the data choose
where they should lie.

\section{Numerical results}
\begin{table*}[t]
\renewcommand{\tabcolsep}{1.1pc} 
\renewcommand{\arraystretch}{1.2} 
\begin{tabular}{|l|ccc|ccc|}
\hline
$P$ &
$\chi^2/$d.o.f.&$\chi^2_{\mathrm{CMD2}}$&$\chi^2_{\mathrm{NA7}}$
&$10^{10} a_\rho$&$10^{10} a_{2 M_K}$ & $\langle r^2 \rangle (\mbox{fm}^2)$ \\ 
\hline
0 & 84.9/83& 43.6  & 43.7 & $420.1\pm2.1$ & $489.5\pm2.2$&
$0.4254\pm0.0020$ \\ 
5 & 78.4/82& 35.9  & 42.6 & $423.8\pm2.6$ & $494.1\pm2.7$&
$0.4300\pm0.0024$ \\ 
6 & 78.1/81& 36.0  & 42.2 & $424.4\pm2.8$ & $494.7\pm2.9$&
$0.4339\pm0.0051$ \\
7 & 73.5/80& 31.7  & 42.2 & $423.4\pm2.9$ & $493.2\pm3.0$&
$0.4350\pm0.0051$\\ 
8 & 73.5/79& 31.6  & 42.2 & $423.5\pm5.7$ & $493.4\pm7.4$&
$0.4347\pm0.0052$ \\
\hline\end{tabular}
\caption{\label{tab:CMD2} Numerical results for fits to CMD-2
  \protect\cite{CMD2} and (spacelike) NA7 data \protect\cite{NA7}. The
  CMD-2 data used here do not contain vacuum polarization effects nor final
  state radiation. The errors given are purely statistical.}
\end{table*}
Our representation of the vector form factor involves three functions, each
of which has free parameters which we pin down by
fitting data. The free parameters at our disposal are the following:
\begin{itemize}
\item in $G_1$ we have two free parameters, the value of the phase
  $\delta_1^1$ at 0.8 and at 1.15 GeV;
\item in $G_2$ we have $P-4$ free parameters, where $P$ is the degree of
  the polynomial in the conformal variable $z$. We will vary $P$ within a
  reasonable range of values, and see how the results depend on it;
\item in $G_\omega$ we have in principle three free parameters,
  $\epsilon_\omega$, $M_\omega$ and $\Gamma_\omega$ -- however, since the
  mass and width of the $\omega$ are rather well known from other
  experiments we fix them at the PDG values. We allow, however, the
  energy calibration (to which $G_\omega$ is very sensitive) to shift
  within the estimated experimental systematic uncertainty.
\end{itemize}
All in all we have $4+(P-4)$ free parameters, depending on the degree of
the polynomial used to describe inelastic effects.

An example of our numerical results obtained by fitting the most recent
CMD-2 data and the spacelike NA7 data \cite{NA7} for values of $P$ between 0
and 8 is given in Table~\ref{tab:CMD2}. The quantities denoted by $a_\rho$
and $a_{2 M_K}$ correspond to the integral (\ref{eq:amuhvp}) cut off at
0.81 GeV and $2M_K$ respectively, whereas $\langle r^2 \rangle$ is the
square of the pion charge radius. It is worth stressing that the good fit
obtained with $P=0$ shows how well the two--pion intermediate state alone
describes the behaviour of the form factor both in the timelike and the
spacelike region. The numerical results obtained in the first line can be
considered as a theoretical prediction of $a_\rho$ based on the calculation
of the $\pi \pi$ phase shift done in \cite{CGL} (of course the results in
that paper is also based on some experimental input). It is clear, however,
that at the level of precision needed for \amuhad\ this prediction is not
sufficiently precise and we must explicitly account for the inelastic
effects encoded in $G_2$.

\begin{table*}[t]
\renewcommand{\tabcolsep}{0.94pc} 
\renewcommand{\arraystretch}{1.1} 
\begin{tabular}{ccccccc}
\hline
$10^{10} a_\rho$ & $10^{10} a_{2 M_K}$ & $E_{max}^{e}/E_{max}^{\tau}$(GeV) &
$\chi^2_{\mbox{\tiny CMD2}}$  
& $\chi^2_{\mbox{\tiny OLYA}}$&$\chi^2_{\mbox{\tiny ALEPH}}$
&$\chi^2_{\mbox{\tiny CLEO}}$ \\ 
\hline
421.5      & --- & .81/---   & 20.3(27) & 14.1(26) &  ---     &  ---     \\
431.0      & --- & ---/.81   &  ---     &  ---     &  8.1(10) & 16.3(21) \\
427.2& --- & .81/.81 & 21.1(27) & 22.4(25) & 11.0(10) & 19.7(21) \\
\hline
427.4&497.7& .97/.81 & 38.2(43) & 27.3(42) & 12.0(10) & 20.1(21) \\
427.7&501.0& .81/.97 & 22.0(27) & 23.9(25) & 12.3(16) & 31.3(28) \\
\hline
\end{tabular}
\caption{\label{tab:both} Results of simultaneous fit to $e^+ e^-$ and
  $\tau$ data in different energy regions. The third column gives the
  maximal energy up to which the $e^+ e^-$ or $\tau$ data are fitted. The
  first two rows concern fits done to either $e^+e^-$ or $\tau$ data. The
  number in brackets next to the $\chi^2$ value gives the number of data
  points fitted.}
\end{table*}
If we switch on the function $G_2$ and allow for one free parameter in it
($P=5$) we observe a sizeable improvement in the $\chi^2$, mainly in the
part which comes from CMD-2 data -- the addition of a further parameter
($P=6$) slightly improves the fit to the NA7 data. With $P=7$ we have
again a sizeable decrease in $\chi^2_{\mbox{\tiny CMD2}}$, whereas going
to $P=8$ brings no improvement anymore but leads to a substantial increase
in the errors on all calculated quantities. Going to even higher values of
$P$ does not make sense anymore, and it is
reasonable to choose as final result the one with $P=7$, while the
variation of the result with $P$ will be included in the final uncertainty,
e.g. by taking the difference among the $P=6$ and $P=7$ results as
theoretical uncertainty. These numbers are preliminary and given only for
illustration purposes: the most important point to stress concerns the
uncertainties, rather than the central values. A comparison to results
obtained with the trapezoidal rule, like e.g. the recent update of
Jegerlehner\footnote{The difference in the central value is mainly due to
  final state radiation, included in (\protect\ref{eq:fred}) but not in
  Tab.~\protect\ref{tab:CMD2}.}  
\cite{Jegerlehner} 
\be
\label{eq:fred}
a_\rho = 429.02 \pm 4.95\, \mbox{ (stat.)}
\ee
shows that the reduction in the purely statistical error is substantial.
Final results will be given in a forthcoming publication
\cite{forthcoming}. We plan to extend the analysis to
other sets of data, like older $e^+ e^-$ data (e.g. \cite{OLYA}), but also
data on the weak vector form factor from $\tau$ decays \cite{ALEPH,CLEO}.

As is well known, the latter sets of data show a systematic deviation from
the $e^+ e^-$ ones, even after correction for known isospin violating
effects \cite{CEN}. One should stress that the calculation of these
effects, although done on a sound theoretical basis, cannot be improved
systematically: it is difficult to estimate the final theoretical
uncertainty in the calculation of these corrections, and it is therefore
not possible to exclude that the remaining discrepancy between $e^+e^-$ and
$\tau$ data is actually due to an unaccounted isospin violating
effect. Indeed this possibility (in particular a difference in mass and
width between charged and neutral $\rho$ mesons) has been discussed at this
Workshop \cite{theseproc} and in a recent publication \cite{Jrhoiso}. 

One of the striking features of this discrepancy is that it has a peculiar
energy dependence: as observed in \cite{DEHZ-2}, the two sets of data are
in good agreement below about 0.8 GeV, but show a marked difference from
that energy on.  It is in fact possible to make a good fit to both sets of
data if one limits oneself to the energy region below 0.8 GeV, as
illustrated in Tab.~\ref{tab:both}. The first two rows give the results of
the fits to either the $e^+e^-$ or the $\tau$ data: the outcome for
$a_\rho$ shows a discrepancy of about 10 units, which is certainly larger
than the error one would like to achieve in this determination. In the
third row, however, it is shown that if one fits simultaneously all data
sets one can still obtain a good $\chi^2$ for all data sets, and a value
for $a_\rho$ which sits between the $e^+ e^-$ and $\tau$ value, somewhat
closer to the latter.  The last two rows show that if one extends the fit
region higher up only for one of the two sets of data, the result for
$a_\rho$ remains stable, whereas the discrepancy in the value of $a_{2
M_K}$ is reduced to about three units.

\section{Conclusions}
I have discussed a method of calculation of the low energy contribution to
\amuhad\ which relies as much as possible on theory -- in the form of
analyticity, unitarity and chiral symmetry. These properties do constrain
the vector form factor of the pion below 1 GeV quite strongly and can be
used to safely interpolate between data. In particular, it is sufficient to
have very precise data in the $\rho$ region to pin down the free parameters
in the form factor and make a controlled extrapolation down to
threshold. In this manner one can sizeably reduce the error in \amuhad\
coming from this energy region, where data are scarce and with larger
errors. I have illustrated this with a few numerical examples and
stressed, in particular, that because of the agreement between $e^+ e^-$
and $\tau$ data below 0.8 GeV, one can give a stable prediction for the
region below this energy. The analysis is still in progress, and final
results will be given in a forthcoming publication \cite{forthcoming}.

In this Workshop we have heard of plans or ongoing efforts for measuring
the pion vector form factor in several different ways, which means that in
a few years time there will be several new sets of data to which this
method can be applied. It will be extremely interesting to look at the
picture that will emerge from these.

\section*{Acknowledgments}
It is a pleasure to thank Marco Incagli and Graziano Venanzoni for their
kind invitation and perfect organization of a short but very intense
Workshop. The work presented here is being done in collaboration with
Irinel Caprini, Heiri Leutwyler and Fred Jegerlehner, whom I warmly thank
-- Heiri Leutwyler also for a careful reading of this manuscript.  Thanks
to Simon Eidelman for useful informations about the CMD-2 data.

\end{document}